\documentclass[useAMS,usenatbib]{mn2e}

\usepackage[pdftex,pdfpagemode={UseOutlines},bookmarks,bookmarksopen,colorlinks,linkcolor={blue},citecolor={blue},urlcolor={red}]{hyperref}
\usepackage{graphicx}	% Including figure files
\usepackage{amsmath}	% Advanced maths commands
\usepackage{amssymb}	% Extra maths symbols
\usepackage{multicol}        % Multi-column entries in tables
\usepackage{bm}		% Bold maths symbols, including upright Greek
\usepackage{pdflscape}	% Landscape pages
\usepackage{caption}
\usepackage{subcaption}
\usepackage{natbib}
\usepackage{longtable,lscape}
\usepackage{rotating}
\usepackage{multirow,multicol}

\usepackage[space]{grffile}  %for including graphics with spaces in the path names

\newcommand{\mnras}{MNRAS}

\newcommand{\aap}{A\&A}
\newcommand{\apj}{ApJ}
\newcommand{\aj}{AJ}
\newcommand{\apjs}{ApJS}
\newcommand{\apjl}{ApJL}

\newcommand{\out}{_\mathrm{out}}
\renewcommand{\in}{_\mathrm{in}}
\def\ba{\begin{align}}
\def\ea{\end{align}}
\usepackage[T1]{fontenc}
\usepackage{ae,aecompl}

\usepackage{newtxtext,newtxmath}
\title[Stability and collisions of triples]{On the stability and collisions in triple stellar systems}

\author[He \& Petrovich]{Matthias Y. He$^{2,1}$\thanks{Contact e-mail: \href{mailto:matthias.he@mail.utoronto.ca}{matthias.he@mail.utoronto.ca}} \&
Cristobal Petrovich$^{1,3}$\\
$^{1}$Canadian Institute for Theoretical Astrophysics, University of Toronto, Ontario, M5S 3H8, Canada\\
$^{2}$Department of Astronomy \& Astrophysics, Pennsylvania State University, University Park, PA 16802, US\\
$^{3}$Centre for Planetary Sciences, Department of Physical \& 
Environmental Sciences, University of Toronto at Scarborough, Toronto, 
Ontario M1C 1A4, Canada
}

\date{Accepted ?. Received ?; in original form ?}

\pubyear{2017}

\begin{document}
\label{firstpage}
\pagerange{\pageref{firstpage}--\pageref{lastpage}}
\maketitle

\begin{abstract}
A significant fraction of main sequence stars are part of a triple system. 
We study the long-term stability and dynamical outcomes of triple stellar systems using a large number of long-term direct N-body  integrations with relativistic precession.
We find that the previously proposed stability criteria by \citealt{EK1995}  and \citealt{MA2001} predict the stability against ejections reasonably well for a wide range of parameters.
Assuming that the triple stellar systems follow orbital and mass distributions from FGK binary stars in the field, 
we find that in $\sim1\%$ and $\sim0.5\%$ of the triple systems lead to a direct head-on collision
(impact velocity $\sim$ escape velocity) between main sequence (MS) stars and between a MS star and a stellar-mass compact object, respectively. 
We conclude that triple interactions are the dominant channel for direct collisions involving a MS star in the field with a rate of one event every $\sim100$ years in the Milky Way. 
We estimate that the fraction of triple systems that
forms short-period binaries is up to $\sim23\%$ with only up to
$\sim13\%$ being the result of three-body interactions with tidal dissipation,
which is consistent with previous work using a secular code.
\end{abstract}

\begin{keywords}
stars: kinematics and dynamics -- binaries: close -- blue stragglers -- white dwarfs
\end{keywords}

%%%%%%%%%%%%%%%%%%%%%%%%%%%%%%%%%%%%%%%%%%%%%%%%%%

%%%%%%%%%%%%%%%%% BODY OF PAPER %%%%%%%%%%%%%%%%%%

\section{Introduction}

Nearly  $20\%$ of the Solar-type stars in the Solar neighbourhood
are in triple stellar systems \citep{DM1991,raghavan2010} and similar or higher fraction is observed for more massive stars (e.g., \citealt{rosa12,Sana12}). 
Moreover, it has been shown that the scatterings of binary-binary, triple-single, and triple-binary systems can produce compact triple systems with configurations subject to strong Lidov-Kozai (LK) oscillations, which can lead to unstable outcomes \citep{AT15}.
Understanding the stability and dynamical fates of these systems have important consequences for gaining insight on their initial assembly, the underlying stellar orbital  architectures, and transient
phenomena associated with close encounters between stars.

Various criteria for the stability of triple stellar systems
have been proposed (see the review by \citealt{G2008} for an extensive list). 
In the general three-body set up (arbitrary masses and orbital configurations), 
these criteria have relied either on direct N-body integrations
(e.g., \citealt{H72,EK1995}) or heuristic methods (e.g., \citealt{MA1999,MA2001}).
Although these criteria have proven useful in many astrophysical applications, their performance using direct long-term N-body integrations has not been properly tested.
In particular, \citet{petrovich2015} showed that two popular criteria by \citet{EK1995} and \citet{MA2001} do not perform very well in the planetary limit (a star orbited by two planets) when the systems are evolved for $10^8$ orbits of the inner body. However, these criteria were proposed in the context of triple stellar systems and, to our knowledge, a numerical study of their performance in this regime is still missing.

Another limitation for astrophysical applications is that 
these stability criteria treat the three bodies as point
masses and ignore forces other than the Keplerian interactions. 
This simplification does not allow for
fates of triple systems other than ejections and long-lived systems, and ignore
possible outcomes when bodies approach each other very closely such as:
mergers, tidal disruptions, and formations of short-period binaries. In particular, short-period binaries can result from the tidal circularization of a binary orbit if a distant third companion induces long term eccentricity excitations (\citealt{MS79}).

In triple stellar systems, these fates have not been studied using long-term
direct N-body integrations because these are generally too expensive to 
capture the relevant dynamical effects that can happen after many 
orbits.
Instead, most previous studies\footnote{Two exceptions are
\citet{VK06} and
\citet{ACR16} who used direct 
N-body integrations of triple systems in globular clusters or short-term
integrations. The latter study evolved 
each triple system for a time equal to the timescale over which they will be 
disrupted through gravitational encounters with other stars.} 
have used the much faster secular codes,
which average all interactions over the inner and outer orbital periods 
(e.g., \mbox{\citealt{EK2006}, \citealt{FT2007}}, \mbox{\citealt{PF09}, \citealt{NF2014}}, \citealt{MK17}).
%\citealt{EK2006,FT2007,PF09,NF2014,MK17}
%\citealt{EK2006}, \citealt{FT2007}, \citealt{PF09}, \citealt{NF2014}, \citealt{MK17}
Although this averaging procedure has proved to be quite useful and
has demonstrated the importance of three-body dynamics at 
producing different fates after close encounters, it 
has a few shortcomings:
\begin{itemize}
\item the triples need to be initialized in stable and
non-crossing orbits ($a\in < a\out [1-e\out]$). For instance,
\citet{FT2007,NF2014} discarded $\sim40\%$ of the randomly 
drawn systems because these were assumed to be 
dynamically unstable in short timescales  given the
stability criterion by \citet{MA2001};
\item processes occurring in dynamical timescales such as
tidal captures and head-on collisions are not properly captured;
\item the most up-to-date study by \citet{NF2014} 
approximates the three-body  gravitational interactions
in powers of the semi-major axis ratio up to the octupole-level
(the Eccentric Kozai-Lidov mechanism, \citealt{naoz2016})
and restrict their analysis to 
$\epsilon_{\rm oct}\equiv a\in e\out/[a\out(1-e\out^2)]<0.1$;
\item the dynamics from the double-orbit averaging 
has recently been shown to poorly describe 
the rate of close approaches (orbits with $1-e\in\ll1$) under
some circumstances.
First, for massive enough perturbers ($m_1\sim m_3$)
it can over-predict the rate of close approaches 
compared to direct N-body integrations even though
$\epsilon_{\rm oct}\sim0.1-0.01$ \citep{LKD2016}.
Second, for systems in mildly hierarchical orbits
$a\out(1-e\out)/a\in\sim3-10$, it can under-predict the
rate of close approaches compared to direct N-body integrations
because the secular approximation breaks down and significant
changes in the orbit's angular moment can occur in dynamical
timescales
\citep{AP12,KD2012,ASTA2014,AMM2014}.
\end{itemize}

In this paper, we test two widely-used criteria by proposed by 
\citet{EK1995} and \citet{MA2001}, for a wide 
distribution of stellar triple systems using intensive N-body 
simulations. 
We track the close approaches between the 
three bodies and include
the effect of apsidal precession due to general relativity.
This approach allows us to estimate the rate of collisions
in a scale-free fashion (independent of the stellar radius)
and we apply it to a variety of triple systems involving main
sequence stars as well as compact objects.

\section{Previous Stability Criteria}

\citet{EK1995} used N-body integrations and handled the eccentricities and mutual inclinations separately, while defining a system to be $n$-stable if it preserves the initial hierarchy of semi-major axes and does not involve any ejections for $10^n$ orbits. They derived an analytical stability boundary for stellar triples based on integrations with $n = 2$ for the outer binary (i.e., $10^2$ outer orbital periods), as given in Equation (\ref{eq:EK}), although they looked at longer evolutions for certain systems.

\citet{MA1999} derived a semi-analytic expression for stability against ejections of the outer body by likening this scenario to stability against chaotic energy exchange in the binary-tides problem. They initially restricted their analyses to the prograde, coplanar case, although a small heuristic factor (the $[1 - 0.3i_{m}/180^{\circ}]$ dependence) was added in order to generalize to mutually inclined orbits, as written in Equation (\ref{eq:MA}) \citep{MA2001,A2004}.

The two stability criteria are re-written here, in a manner following that of \citet{petrovich2015}, where we use 
\begin{align}
r_{\rm ap} \equiv \frac{a_{\rm out}(1-e_{\rm out})}{a_{\rm in}(1+e_{\rm in})}
\end{align}
as the orbit separation parameter ($a_{\rm in}$ and $a_{\rm out}$ refer to the semi-major axes of the inner and outer bodies respectively, while $e_{\rm in}$ and $e_{\rm out}$ refer to the eccentricities of the inner and outer body orbits, respectively) and write the conditions for stability in the form of $r_{\rm ap} > Y$, with $Y_{\rm EK}$ and $Y_{\rm MA}$ to denote the results from the
\citet{EK1995} and \citet{MA2001}, respectively, as:
\begin{align}
Y_{\rm EK} \equiv{}& 1 + \frac{3.7}{q_{\rm out}^{1/3}} - \frac{2.2}{1 + q_{\rm out}^{1/3}} + \frac{1.4}{q_{\rm in}^{1/3}}\frac{q_{\rm out}^{1/3} - 1}{1 + q_{\rm out}^{1/3}} 
\label{eq:EK}\\
\begin{split}
Y_{\rm MA} \equiv{}& 2.8\frac{1}{1 + e_{\rm in}}\bigg[{(1 + \mu_{\rm out})}\frac{1 + e_{\rm out}}{(1 - e_{\rm out})^{1/2}}\bigg]^{2/5} \\
 & (1 - 0.3i_{m}/180^{\circ}).
\label{eq:MA}
\end{split}
\end{align}
Here, $q_{\rm in} = m_1/m_2$ and $q_{\rm out} = (m_1+m_2)/m_3$ are the inverse mass ratios of the inner and outer bodies respectively to the bodies within, and $\mu_{\rm out} = m_3/(m_1+m_2)$ is the mass ratio of the outer body to the inner bodies (1, 2, 3 denote the bodies from innermost to outermost). We refer to the conditions for stability involving Equations (\ref{eq:EK}) and (\ref{eq:MA}), i.e. 
$r_{\rm ap} > Y_{\rm EK}$ and $r_{\rm ap} > Y_{\rm MA}$, hereafter as EK95 and MA01 respectively.

\section{Our Numerically Simulated Sample}

To simulate the triple systems over a long timescale, we ran N-body integrations using the IAS15 integrator (\citealt{RS2015}) of REBOUND (\citealt{RL2012}). This integrator uses an adaptive time-stepping algorithm to keep systematic errors below machine precision, and is well suited to handle close encounters and highly eccentric orbits (\citealt{RS2015}). 

We use modules from the additional REBOUNDx package to include effects of general relativity (GR) and use the ``grfull" implementation detailed in (Tamayo et al., in prep\footnote{More details can found in the following link: \mbox{\url{http://reboundx.readthedocs.io/en/latest/effects.html}}}), which advances the pericentre according to the point-mass acceleration from 
\cite{newhall1983} (Equation 1 therein),
valid for separations $\Delta r\gg2G(m_i+m_j)/c^2$. 
For $m_1+m_2=1M_\odot$, this distance corresponds to $2.96\times10^5 \mbox{ cm}\simeq2\times10^{-8}\mbox{ AU}$. 
We limit our analysis to separations of $\Delta r>10^{-6}$ AU for which our approximation should work well. 

We do not include tidal forces nor rotations of the stars in our simulations. This simplification has the advantage that our results are independent of the radii of the bodies and can be applied to different systems containing main sequence (MS) stars, white dwarfs (WD), neutron stars (NS), and black holes (BH). Since  we are limited to minimum separations much larger than the NS radius
($\Delta r\gg 1R_{\rm NS}\sim10^{-7}$ AU), 
the systems for which we can track close encounters need to either 
have a main sequence star or a white dwarf.

\subsection{Distribution of Initial Parameters}

We draw a sample of 5000 systems in a similar manner as \citet{FT2007}, which is based on the observed populations in \citet{DM1991}. The central mass is set to $m_1 = 1M_{\odot}$; the inner body mass $m_2$ is chosen by drawing the ratio $\mu_{\rm in} = m_2/m_1$ from a Gaussian distribution with a mean of 0.23 and standard deviation of 0.42, and then the outer body mass $m_3$ is chosen in the same manner by drawing $\mu_{\rm out} = m_3/(m_1+m_2)$ from the same Gaussian distribution. Negative results are resampled and thus the actual distributions are skewed to slightly larger values. We do not assign any radii to our three bodies, and thus treat them as point sources and do not directly simulate any collisions. Instead, we track various degrees of close encounters which we can then consider what would be collisions or mergers at any scale we choose.

The initial semi-major axes ($a_{\rm in}$, $a_{\rm out}$) are chosen by sampling orbital periods ($P_{\rm in}$, $P_{\rm out}$) from a lognormal distribution with mean of $\log{(P/\text{days})}$ of 4.8 and standard deviation of 2.8. We impose a cutoff of 1 day for the periods by resampling periods shorter than this. The shorter period is then assigned to the inner body and the larger period to the outer body.

We sample the initial eccentricities ($e_{\rm in}$, $e_{\rm out}$) uniformly in $[0,0.9]$, which are the only orbital parameters drawn differently than in \citet{FT2007} who considered a thermal eccentricity distribution without an upper limit. The uniform eccentricity distribution is favored by the observations of binary systems in \cite{raghavan2010}
and triple systems in \citet{TK16}, while the upper limit of 0.9 is arbitrary and only imposed to limit the number of short-lived systems in orbit-crossing trajectories. 
We set the initial inclination of the inner orbit to 0 and draw the initial outer inclination as $\cos{i}$ uniformly in $[-1,1]$ to assume isotropic initial mutual inclinations. 
The other angles and arguments ($\omega_{\rm in}$, $\Omega_{\rm in}$, $f_{\rm in}$ and similarly for the orbit of the outer body) are sampled uniformly in $[0,2\pi]$.

\subsection{Integration and Stopping Conditions}

Instead of integrating every system that is drawn, we first compute and compare some dynamical timescales to decide which systems are interesting enough to simulate. 

Sufficiently hierarchical systems ($r_{\rm ap}\gg1$) are stable in dynamical timescales and they can only evolve secularly without exchanging energy, but only exchanging angular momentum. This angular momentum exchange can potentially drive the inner orbit's eccentricity to nearly unity by the Lidov-Kozai
mechanism \citep{lidov,kozai}, leading to strong tidal interactions and different possible outcomes (see e.g., \citealt{naoz2016}). 
A necessary, but not sufficient, condition for this to happen is that the characteristic 
 Lidov-Kozai timescale (e.g., \citealt{A15}):
\begin{align}
\tau_{\rm LK} &= \frac{2}{3\pi} \frac{P_{\rm out}^2(1-e_{\rm out}^2)^{3/2}}{P_{\rm in}} \frac{m_1+m_2+m_3}{m_3}
\label{tLK}
\end{align}
overcomes the general relativistic (GR) orbital precession timescale $\tau_{\rm GR}$ (e.g., \citealt{FT2007}):
\begin{align}
\tau_{\rm GR} &= \frac{c^2}{3G^{3/2}}\frac{a_{\rm in}^{5/2}(1-e_{\rm in}^2)}{(m_1+m_2)^{3/2}} \nonumber\\
&\simeq 5.3\mbox{ Myr }\left(\frac{a_{\rm in}(1-e_{\rm in}^2)^{2/5}}{1\mbox{ AU}}\right)^{5/2}
\left(\frac{M_\odot}{m_1+m_2}\right)^{3/2}.
\label{tGR}
\end{align}
We disregard the systems with $\tau_{\rm LK} > \tau_{\rm GR}$ and $r_{\rm ap} > 10$ and classify them as long-term stable.

We set a maximum integration time of $t_{\text{max}} = 10^7 P_{\rm in}$, and only simulate systems satisfying both $\tau_{\rm LK} < \tau_{\rm GR}$ and $\tau_{\rm LK} < t_{\text{max}}$. During each system's integration, we check for close encounters between any bodies and record the first time the system involves a close approach of less than $10^{-1}$AU, $10^{-1.25}$AU, $10^{-1.5}$AU, and so on to $10^{-6}$AU. The integration is stopped when one of the following conditions is met:
\begin{itemize}
\item an ejection occurs, which we define as having the distance $d$ of a body to the centre of mass exceed five times the initial semi-major axes of the outer body, i.e. $d > 5a_{\rm out,i}$\footnote{If one of the bodies satisfies $d > 5a_{\rm out,i}$ it is almost always in an unbound orbit (ejected). Even if it is not ejected, the configuration changed significantly relative to the initial one, so the initial configuration is considered unstable.};
\item the system has evolved for $t_{\text{\rm max}} = 10^7 P_{\rm in}$;
\item a computational wall-time of 48h is reached.
\end{itemize}

Of the 5000 systems, only 3207 were simulated (1688 have $\tau_{\rm LK} > \tau_{\rm GR}$ and 105 have $\tau_{\rm LK} > t_{\text{max}}$). Out of these, 1222 did not finish integrating to $10^7 P_{\rm in}$ (i.e., they only met the last condition above). The majority of these evolved for well over $10^6 P_{\rm in}$ (only 221 systems did not finish integrating to $5\times10^6 P_{\rm in}$); thus in order to improve statistics we reduce our total evolution time to $t_{\text{max}} = 5\times10^6 P_{\rm in}$ when considering the outcomes for each system.

We then classify our systems as stable or unstable based simply on their dynamical outcomes over the integration time: a system is unstable if there is an ejection and stable if there are none. Since we are treating all three bodies as point masses and thus not directly simulating collisions or mergers, all systems fall into these two categories; however, by tracking close encounters we can distinguish which systems would have resulted in mergers or collisions at various scales. Additionally we discard (60) systems where the inner body has an initial pericentre of less than 0.01 AU, i.e. $a_{\rm in,0}(1-e_{\rm in,0}) < 0.01$ AU, in order to prevent systems beginning with very close-approaching configurations to the host star from contributing to our subsequent analyses involving close encounters and collisions of the systems. Our final sample includes 4720 systems.

\section{Results}

\subsection{Performance of stability criteria}

\begin{figure*}
\center
\includegraphics[width=\textwidth]{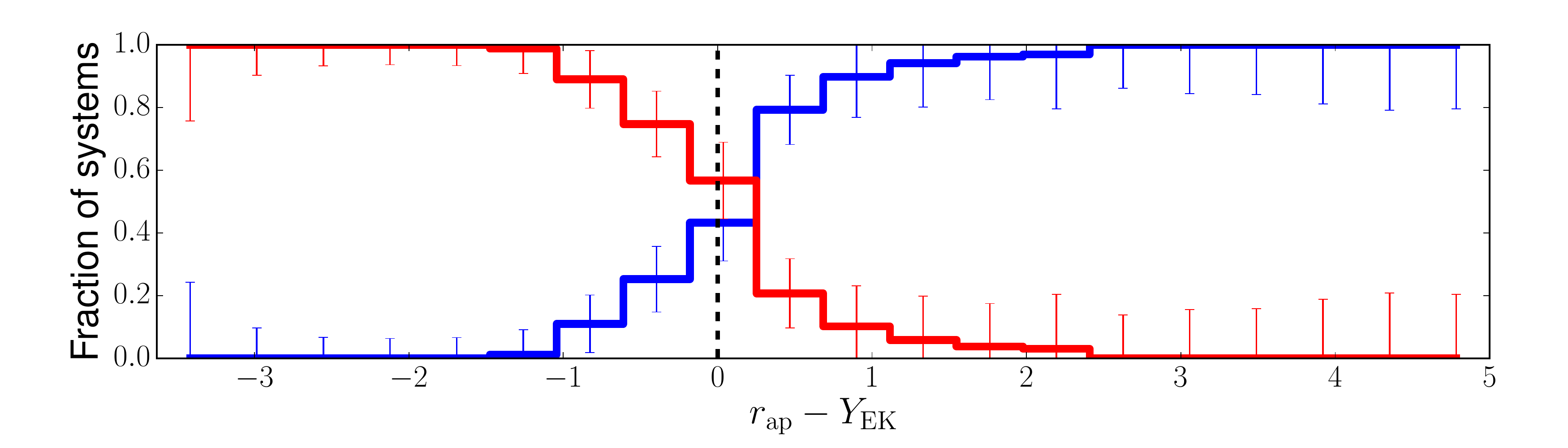}
\includegraphics[width=\textwidth]{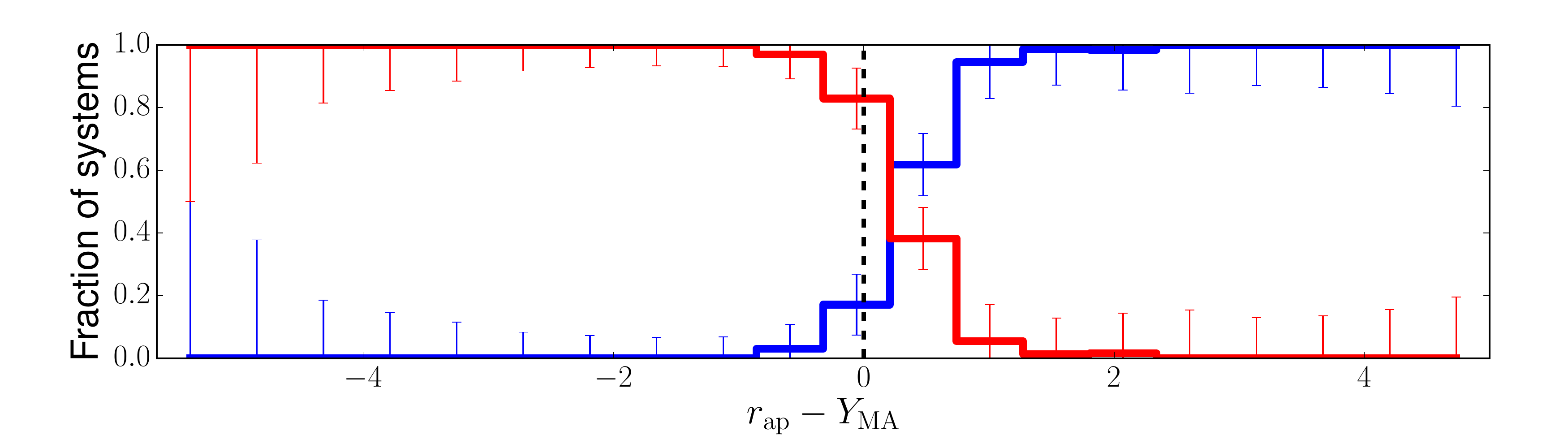}
\caption{Histograms of the relative fractions of stable (blue) and unstable (red) systems as a function of stability conditions, $f = r_{\rm ap} - Y_{\rm EK}$ (top) and $f = r_{\rm ap} - Y_{\rm MA}$ (bottom). Systems involving close encounters at the $\Delta{r} < 0.01$ AU scale are not included here. The black dashed vertical line marks the stability boundary $f = 0$, above which systems are classified as stable and below which systems are classified as unstable according to the criteria. We use 20 uniform bins between the minimum value of $f$ and $f = 5$; systems with $f > 5$ are not shown (they are all extremely stable).}
\label{fig:stabilitycriteria}
\end{figure*}

In Figure \ref{fig:stabilitycriteria}, we plot the relative fractions of stable systems and their complement, unstable systems, as a function of the stability criterion $f \equiv r_{\rm ap} - Y$ where $Y = Y_{\rm EK}$ (given by Equation \ref{eq:EK}) in the top plot and $Y = Y_{\rm MA}$ (given by Equation \ref{eq:MA}) in the bottom plot. Since the conditions for stability are written in the form of $r_{\rm ap} > Y$, the parameter $f$ quantifies the degree of stability and defines a boundary at $f = 0$, with $f > 0$ predicting stability and $f < 0$ predicting instability over a long time scale. Systems involving close encounters of less than $10^{-2}$ AU $\sim2R_\odot$ and systems that were not simulated due to obeying $\tau_{\rm LK} > \tau_{\rm GR}$ or $\tau_{\rm LK} > t_{\text{max}}$\footnote{The systems with $\tau_{\rm LK} > \min\{\tau_{\rm GR},t_{\text{max}}\}$ have $f\gg1$ and are all expected to be stable.}
are not included here; a total of 2790 systems are plotted. The highly stable tail (upper end of the x-axes) is also truncated at $f = 5$ for clarity.

We see that both the EK95 and MA01 criteria perform quite well for the range of systems we explored.
By defining the completeness of the classifications (stable or unstable) as the ratio of the number of correctly identified systems satisfying $f > 0$ (or $f < 0$) to the total number of stable (or unstable) systems, we quantify the performances of the criteria as follows: the EK95 criteria provides a completeness of about 96.6\% (1520/1573) for stable systems and 95.9\% (1167/1217) for unstable systems; the MA01 criteria provides a completeness of about 98.9\% (1555/1573) for stable systems and 93.6\% (1139/1217) for unstable systems.

We caution that the completeness depends on the distribution of initial conditions and if we had populated the region around $f=0$ more densely relative to regions with $|f|\gg1$, the completeness would likely decrease.
In particular, if we restrict to the initial conditions in the range $f=[-3,3]$ we get a completeness of 86.8\% (95.7\%) and 95.2\% (92.6\%) for stable (unstable) systems using EK95 and MA01, respectively.
Even in these more unfavorable limits, the completenesses are still relatively high, especially using the MA01 criteria. Thus, we consider that the overall separation of the classes is good enough for practical purposes (observational works and theoretical population synthesis models). The mixed states occur mostly at $|f|\lesssim0.05$.

It is important to note that the orbit separation parameter $r_{\rm ap}$ itself is already a very good stability indicator. There is a clear constant value of $r_{\rm ap}$ which seems to form the general boundary between stable and unstable systems. To test how well the EK95 and MA01 criteria perform over the $r_{\rm ap}$ parameter alone ($f = r_{\rm ap} - Y$ where $Y$ is a constant), we compute the stable and unstable completeness for $Y = 3$ and $Y = 2.5$. Using the same systems included in Figure \ref{fig:stabilitycriteria} (2790 systems), the $r_{\rm ap} = Y = 3$ ($f = 0$) boundary results in a completeness of 95.9\% and 97.5\% for stable and unstable systems, respectively. Restricting the domain to systems with $f=[-3,3]$, the completenesses are 84.1\% and 97.5\%. The $r_{\rm ap} = Y = 2.5$ ($f = 0$) boundary results in a completeness of 98.2\% and 93.4\% for stable and unstable systems, respectively (for the 2790 systems), while taking the subset of systems with $f=[-3,3]$ gives 92.1\% and 93.4\%. As expected, shifting to a lower constant $Y$ improves the completeness of stable systems while trading off the completeness of unstable systems, as a lower $Y$ is equivalent to having a less conservative criteria for stability. 

Interestingly, we note that the simple conservative criterion $r_{\rm ap}>3$ above is 
consistent with the observations of triple systems in \citet{T08}
who finds that these systems satisfy $P\out/P\in\gtrsim5$: 
the criterion $r_{\rm ap}>3$ implies  
 that $a\out/a\in>3$ ($e\in=e\out=0$), which for
equal-mass stars implies that $P\out/P\in>3^{3/2}\times (2/3)^{1/2}=4.24$.
As previously pointed out by \citet{petrovich2015} the constant $r_{\rm ap}$ criterion also works well in the planetary regime ($m_2,m_3\ll m_1$), but for a smaller
value as expected: $r_{\rm ap}=1.83$ instead of $r_{\rm ap}=2.5-3$ for triple stars.

In summary, based on the performance in the classification of our 
simulated systems we conclude that the MA01 criterion 
performs very well with  completeness of $\sim95\%$. 
The criterion by EK95 performs slightly worse than MA01 and it is
comparable to the simple constant criterion $r_{\rm ap}=2.5$.

\subsection{Close encounters and ejections}

\begin{figure*}
\center
\includegraphics[width=\textwidth]{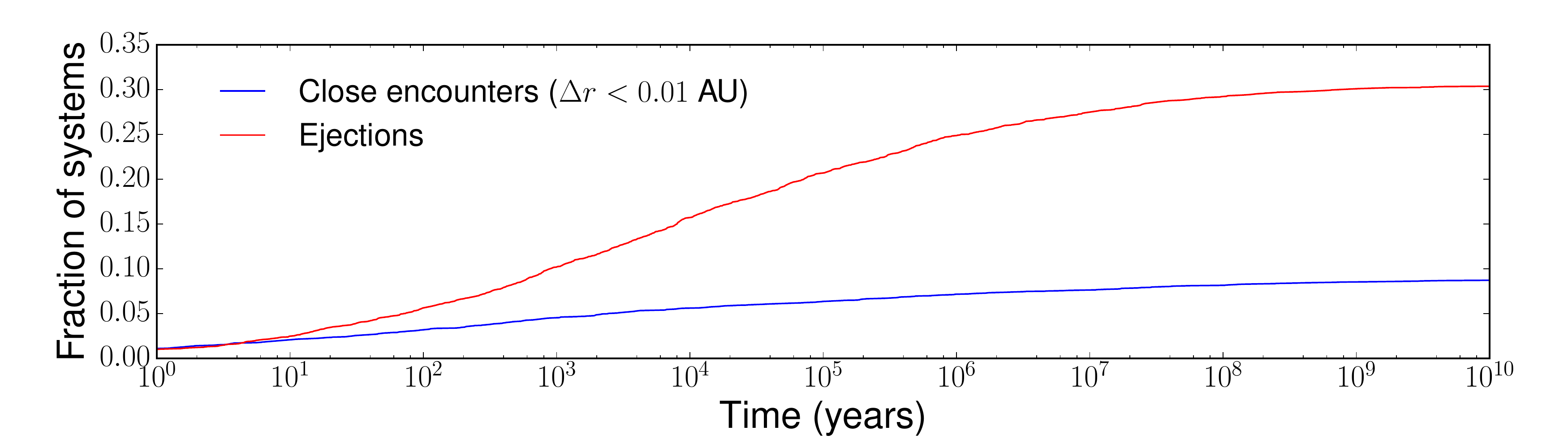}
\includegraphics[width=\textwidth]{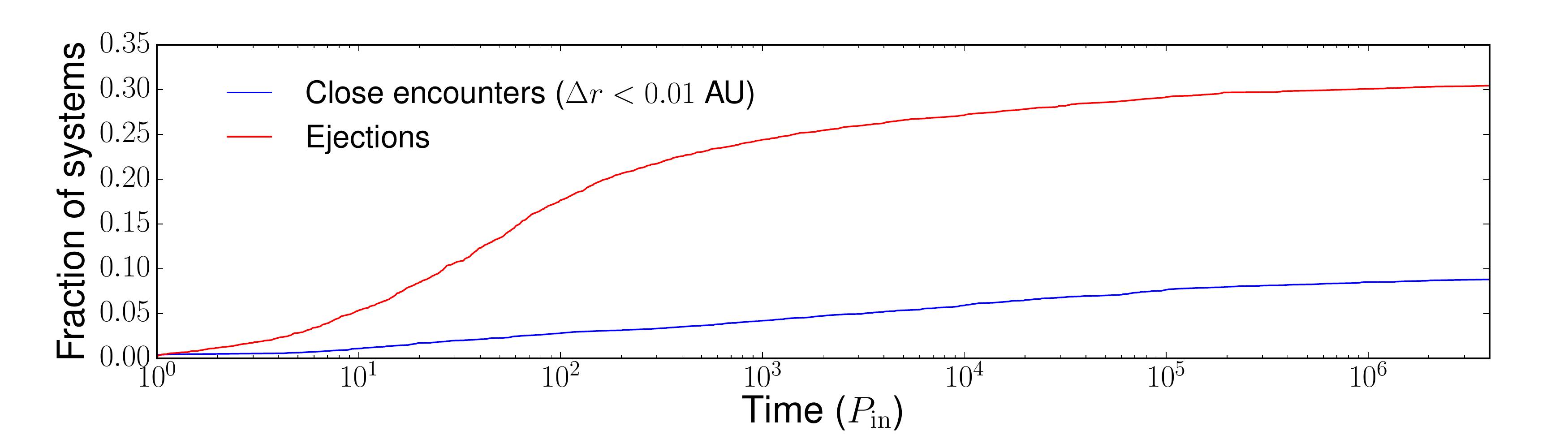}
\caption{Plot of the cumulative fraction of systems involving close encounters of $\Delta{r} < 0.01$ AU (blue) and ejections (red) as a function of time in years (top) and in periods of the inner body, $P_{\rm in}$ (bottom).}
\label{fig:time_ejectmerger}
\end{figure*}

We show plots of the cumulative fractions of systems involving close encounters ($\Delta{r} < 0.01$ AU $\sim2R_\odot$) and ejections over time, in Figure \ref{fig:time_ejectmerger}. The timescales are shown both in units of years and in units of the initial period of the inner body, $P_{\rm in}$. We restrict the x-axes to range from 1 yr to $10^{10}$ yr and from 1 $P_{\rm in}$ to $5\times10^6 P_{\rm in}$, respectively. The number of systems involving ejections and thus becoming unstable increases significantly over time, with a rate (scaled logarithmically to time) that seems to increase until roughly $10^5$ yr or $10^2 P_{\rm in}$. It then levels off horizontally for very large ages, to a value that is just above 30\% (1440/4720) of our total number of systems. 
In turn, the number of systems involving close approaches inside $\Delta{r} = 0.01$ AU increases roughly at a constant rate as a function of $\log P_{\rm in}$, although a smaller number of systems reach this criteria; a final value of 8.54\% (403/4720) represents the total fraction of systems reaching such a close encounter in $5\times10^6 P_{\rm in}$ of evolution, for the distribution of systems we explored.

\begin{figure*}
\center
\includegraphics[width=\textwidth]{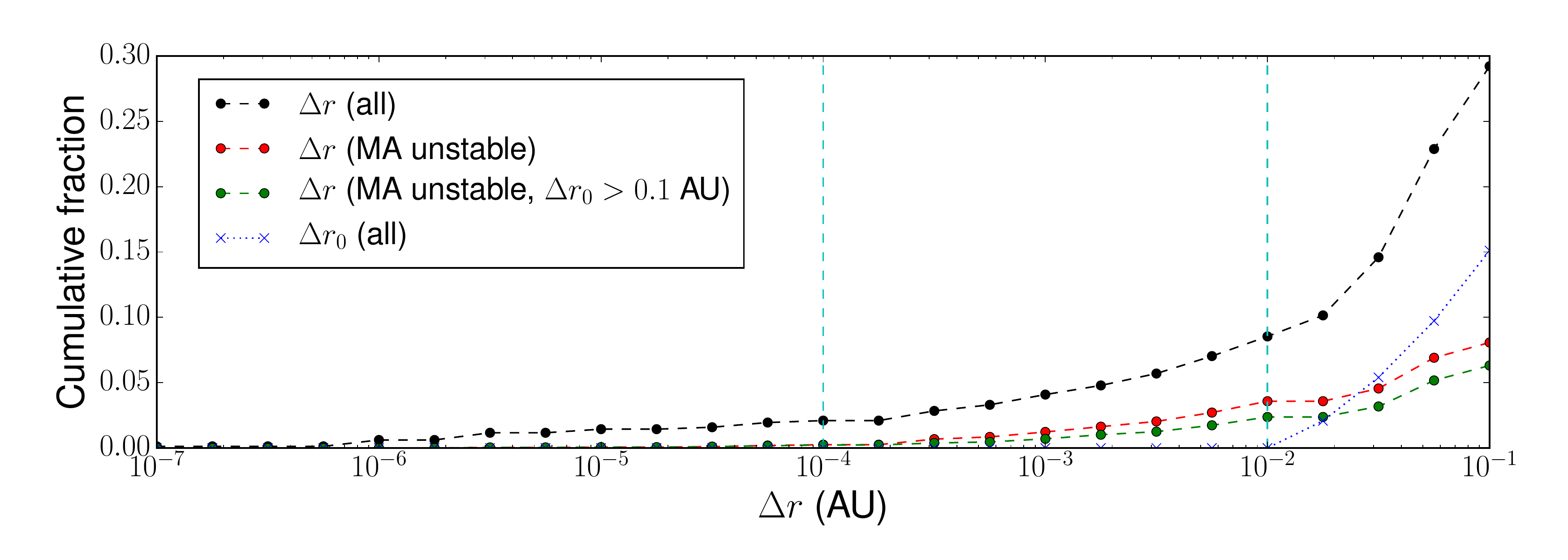}
\caption{Cumulative distribution function for fractions of systems reaching various close encounter thresholds $\Delta{r}$. The black curve includes systems reaching $\Delta{r}$ at any point of their evolution (including the initial distribution, $\Delta{r_0}$). The red curve includes only the systems that are classified as unstable by the Mardling \& Aarseth boundary, $f = r_{\rm ap} - Y_{\rm MA} < 0$, while the green curve is a subset of the red curve, also excluding systems with $\Delta{r_0} = a_{\rm in,0}(1 - e_{\rm in,0}) < 0.1$ AU. The blue curve shows the initial cumulative distribution of the closest approach, the initial pericentre distance $\Delta{r_0} = a_{\rm in,0}(1 - e_{\rm in,0})$.}
\label{fig:mergerCDF}
\end{figure*}

In Figure \ref{fig:mergerCDF}, we plot several curves for the cumulative fractions of systems reaching various close encounter thresholds (as recorded in our integrations). We use $\Delta{r}$ to denote the lowest threshold that any two bodies in the system, relative to each other, surpassed at any time during their orbital evolutions, and $\Delta{r_0} = a_{\rm in,0}(1 - e_{\rm in,0})$, i.e. the initial pericentre of the inner body.  The black curve includes all our systems, while the blue-`x' curve shows the distribution of the initial closest separation, $\Delta{r_0}$. Note that as mentioned previously, we discarded systems with $\Delta{r_0} < 10^{-2}$ AU as they would contaminate the actual close encounters that we tracked. Thus, we get some contribution from the initial distribution to close encounters down until $\Delta{r} = 10^{-2}$ AU, where it cuts to zero. The red and green lines both show the cumulative fractions of systems reaching each $\Delta{r}$ for systems classified as unstable by MA01, while the latter also excludes systems from the blue-`x' curve, i.e. systems with $\Delta{r_0} < 0.1$ AU. Comparing these two curves, it is clear that systems with initially close-in orbits of the inner body ($\Delta{r_0} < 0.1$ AU) contribute to the number of close encounters we have down to almost $10^{-4}$ AU, for MA01 unstable systems. The difference between the black and red plots show that while systems reaching $\Delta{r}$ between $10^{-1}$ and $\sim10^{-4}$ AU come from both stable and unstable (MA01) systems, almost all of the extremely close encounters less than $\Delta{r} \simeq 10^{-4}$ AU are in stable systems with eccentricity-driven close approaches.

\begin{figure*}
\center
\includegraphics[width=\textwidth]{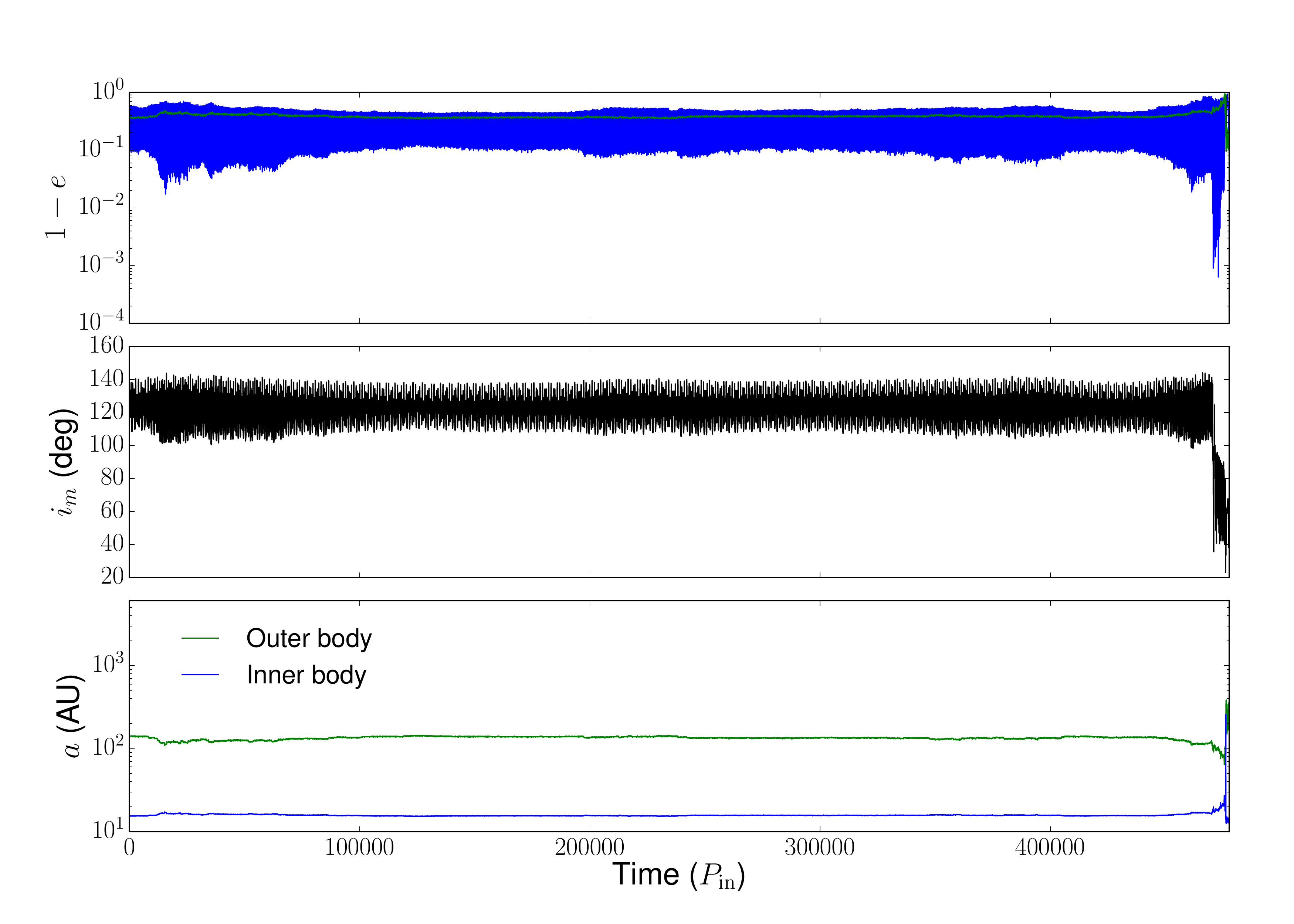}
\caption{Evolution of a system close to the stability boundary ($r_{\rm ap}-Y_{\rm MA}\simeq-0.198$) that resulted in an ejection. The top panel shows the eccentricity oscillations, the middle panel shows the mutual inclination oscillations, and the bottom panel shows the evolution of semi-major axes of the inner and outer bodies over time in $P_{\rm in}$. The inner body was initialized with $m_2 = 0.070 M_\odot$, $a_{\rm in} =  15.324$ AU, $e_{\rm in} = 0.746$, $i_{\rm in} = 0$, $\Omega_{\rm in} = 0$, $\omega_{\rm in} = 1.121$, and $f_{\rm in} = -2.047$ (giving an initial $P_{\rm in} \simeq 58$ yr), while the outer body was initialized with $m_3 = 0.217 M_\odot$, $a_{\rm out} =  143.136$ AU, $e_{\rm out} = 0.646$, $i_{\rm out} = 2.030$, $\Omega_{\rm out} = -3.051$, $\omega_{\rm out} = 1.868$, and $f_{\rm out} = 6.386$. This system satisfies our criteria for a MS-MS clean collision (as defined in section 5.1) near the end of its evolution. The system also experiences a close encounter of $\Delta{r} \simeq 0.001$ AU after $\simeq 4.73\times10^{5} P_{\rm in}$ and an ejection shortly thereafter.}
\label{fig:evolution}
\end{figure*}

In Figure \ref{fig:evolution} we show the evolution of one system
that is initially close to the instability edge with $r_{\rm ap}-Y_{\rm MA}\simeq-0.198$ 
to illustrate the peculiar behavior that can give rise to both collisions
and ejections.
From the upper and middle panels we observe that this system undergoes 
large-amplitude eccentricity and inclination oscillations
on timescales of $P_{\rm out}$, which is similar to the 
chaotic excursions of the pericentre described by 
\citet{AP12,KD2012} when the secular approximation breaks down. However, unlike 
what these authors described, the system is at the stability edge
and it becomes unbound after $\sim4.7\times10^5 P_{\rm in}$. Before doing so, it experiences close encounters as close as $\Delta{r} \simeq 0.001$ AU. It also satisfies our definition of a clean collision between main sequence stars as defined in section 5.1 (but does not constitute a clean collision between a main sequence star and a compact object, as defined in section 5.2). Thus, 
in this example the random diffusion of the pericentre distance is also
accompanied by a slow diffusion in energy.

\subsection{Effect of initial mutual inclination on stability}

\begin{figure*}
\center
\includegraphics[width=\textwidth]{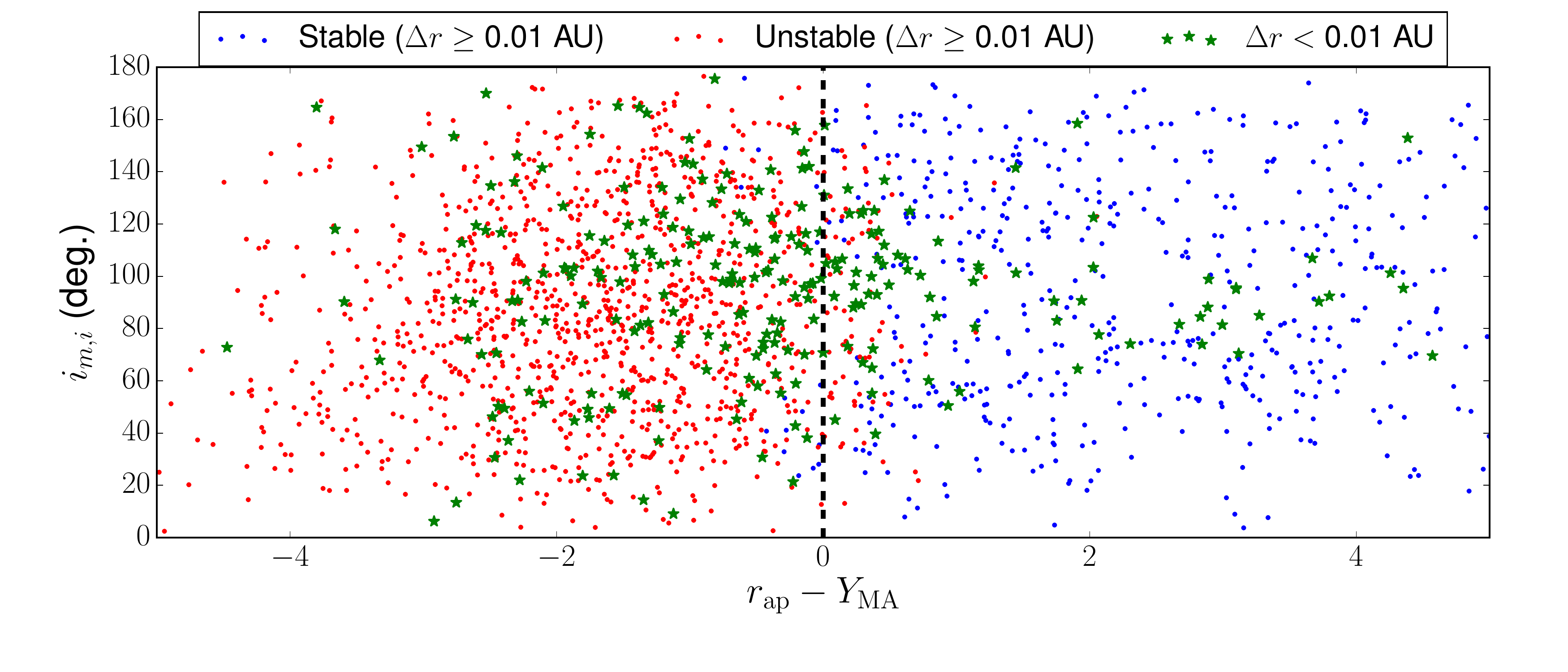}
\caption{Initial mutual inclination $i_{m,0}$ vs. the \citet{MA2001} criteria parameter $f = r_{\rm ap} - Y_{\rm MA}$ scatter plot of our systems, colour-coded by stability outcomes. Green stars `*' indicate systems reaching close encounters less than $\Delta{r} = 0.01$ AU. Blue and red dots denote systems which never involve encounters less than $\Delta{r} = 0.01$ AU and are stable and unstable, respectively. The black dashed vertical line marks the stability boundary $f = 0$.}
\label{fig:incMA}
\end{figure*}

In order to assess the dependence of stability on the initial mutual inclination of the two bodies, we present a scatter plot in Figure \ref{fig:incMA} of $i_{\rm m,i}$ versus $f = r_{\rm ap} - Y_{\rm MA}$ for our simulated systems. We separate these systems into three categories for this plot: systems with $\Delta{r} < 0.01$ AU (green `*'s), and stable (blue) and unstable (red) systems with $\Delta{r} > 0.01$ AU. The unstable and stable systems are scattered left and right, respectively, of the MA01 stability boundary $f = 0$ as we expect from the stability criteria histograms in Figure \ref{fig:stabilitycriteria}, and rather uniformly over a wide range of $i_{m,i}$. However, the systems reaching $\Delta{r} < 0.01$ AU are distributed differently. These close encounters occur in systems classified as stable and unstable by MA01. The green `*'s with $f < 0$ are distributed over a wide range of $i_{\rm m,i}$, although they appear to be slightly more concentrated around $i_{\rm m,i} \sim 90^\circ$, while the green `*'s with $f > 0$ occur almost exclusively around $i_{\rm m,i} \sim 90^\circ$. Note that in the MA01 criteria, there is a linear correction factor involving $i_{\rm m}$ up to 30\%, which \citet{MA2001} included to account for the added stability of inclined systems. We remark that the inclusion of this factor is appropriate, as omitting it caused the systems with high $i_{\rm m,i}$ to shift towards lower $f$ and thus more stable systems (blue) to cross the stability boundary $f = 0$ into the negative region.

%%%%%%%%%%%%%%
\section{Direct collisions and formation of short-period
binaries}

We discuss the possible outcomes that are expected for systems
reaching close encounters such as collisions and the formation of 
short-period binaries.
We remind the reader that we have not included tidal interactions
between the stars, but since we have kept track of the close approaches
we can still estimate the rate at which these outcomes occur.

\subsection{Clean collisions between Main-Sequence stars}

An interesting by-product from our calculations is an estimate of the 
rate of clean head-on collisions between main sequence stars in the field, 
which has not been worked out previously. 
A clean collision avoids a tidal capture and the impact occurs 
at velocities that are similar to the escape velocities of the stars.
The outcomes from these collisions
have been studied using SPH simulations (e.g., \citealt{FB05})
and could lead to observable 
transient events (e.g., \citealt{SR98,ST06}), possibly
similar to that observed in V1309 Scorpii (e.g., \citealt{Tylenda2011}).

\begin{figure*}
\center
\includegraphics[width=\textwidth]{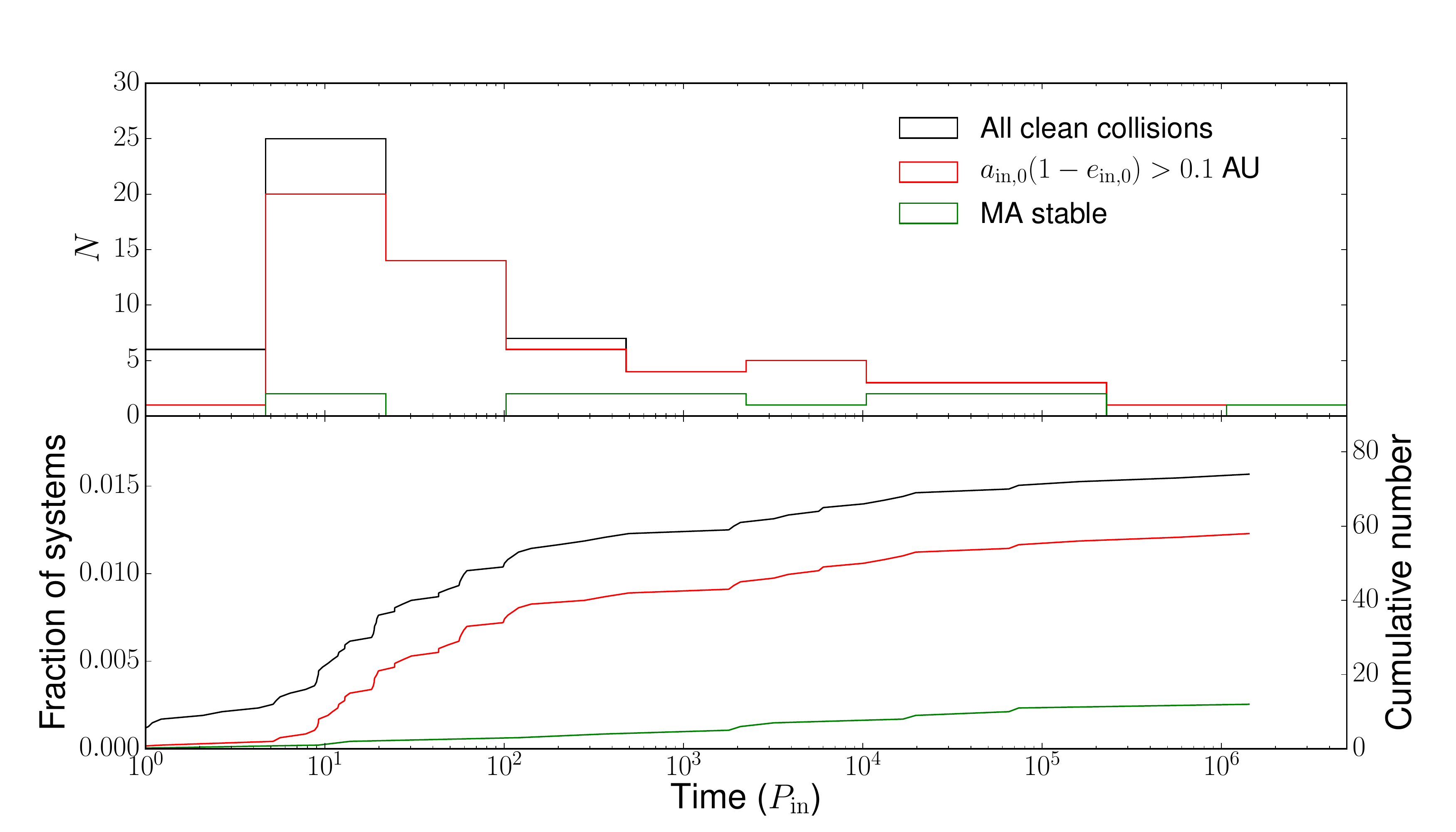}
\caption{Number of clean collisions between MS stars (as defined in section 5.1) in our simulations as a function of time. \textbf{Top:} Histograms of the time taken (in $P_{\rm in}$) for the systems to involve a clean collision at the $\Delta{r} = 0.01$ AU scale, satisfying various conditions. We use 10 bins spaced uniformly in log, from $1 P_{\rm in}$ to $5\times10^6 P_{\rm in}$. \textbf{Bottom:} Cumulative fractions (numbers) of systems involving a clean collision as a function of time (in $P_{\rm in}$) for various systems.}
\label{fig:cleancolMS}
\end{figure*}

We define a clean collision such that $\Delta{r} < R_1+R_2 \simeq 2R_\odot \simeq0.01$ AU at some point, but previous to this event (i.e., previous pericentre passages) it avoided suffering a tidal disruption event or tidal capture. 
Based on the models of dynamic tides by \citet{PT77} and \citet{LO86}
we can estimate the amount of energy deposited in the primary ($m_1=M_\odot$
with polytropic index of 3) at closest approach $\Delta r\gtrsim4R_1$
\begin{align}
\Delta E_{\rm tide} &= \frac{Gm_1^2}{R_1}
\left(\frac{m_2}{m_1}\right)
\left(\frac{R_1}{\Delta r}\right)^6T_2+\mathcal{O}\left(
\left[\frac{R_1}{\Delta r}\right]^8T_3\right),
\end{align}
where only leading-order quadrupolar perturbations are considered (term accompanying $T_2$).
This approximation is reasonable for $\Delta r\gtrsim4R_1$ since $T_3<T_2$ \citep{LO86}. 
The term $T_2$ is a function of the masses
and $R_1/\Delta r$, and for $m_1=m_2$ we have $T_2<0.01$
\citep{LO86}. Thus, for two Solar-type stars the fraction of orbital energy 
that can be dissipated through dynamical tides is
\begin{align}
\frac{\Delta E_{\rm tide}}{E_{\rm orb}}
&< 0.01\left(\frac{a_{\rm in}}{10 \mbox{ AU}}\right)\left(\frac{4R_\odot}{\Delta r}\right)^6,
\end{align}
meaning that each pericentre passage with $\Delta r\gtrsim4R_\odot$ changes
the semi-major axis by $|\Delta a_{\rm in}|/ a_{\rm in} <0.01$
if $a_{\rm in}<10$ AU. In other words, order-unity changes are expected after 
$\sim100$ pericentre passages. Note that $\Delta E_{\rm tide}$ falls-all
much more rapidly than $\Delta r^{-6}$ because $T_2$ also decays rapidly with 
$\Delta r$.

We define a conservative threshold distance of $\Delta{r} =10^{-1.5}\simeq0.031$ AU $\simeq 6.7R_\odot$, and require that the spiraling from $\Delta{r} \simeq 0.03$ to 0.01 AU (i.e. $10^{-1.5}$ to $10^{-2}$ AU in our tracked close encounters) occur in less than one orbit, i.e. $\Delta{t} < 1 P_{\rm in}$, in order to guarantee the avoidance of tidal capture. This threshold
is conservative because the semi-major axis only changes by $|\Delta a_{\rm in}|/ a_{\rm in} \lesssim 10^{-5} \times a_{\rm in}/(10 \mbox{ AU})$ per pericentre passage. 

Our calculations indicate that clean collisions occur in $1.61\%$ (76/4720) of our systems. The system shown in Figure \ref{fig:evolution} is an example of a system that experienced a clean collision. If we limit our results to only include systems with $\Delta{r_0} = a_{\rm in,0}(1 - e_{\rm in,0}) > 0.1$ AU, we still get a clean collision rate of $1.25\%$ (59/4720). If we are even more conservative and only consider stable systems (MA01 stable and avoids ejections over $t_{\rm max}$), the clean collision rate is $\sim0.4\%$ (13/3489). 
If a less conservative threshold for the tidal barrier of $\Delta r=10^{-1.75} $ AU $\simeq3.8 R_\odot$ instead of $\simeq6.7R_\odot$ is used, the number of clean collisions increases from 75 to 123.

In the top panel of Figure \ref{fig:cleancolMS} we plot histograms of the times taken for the systems to result in a clean collision, in terms of the initial period of the inner body $P_{\rm in}$ on a logarithmic scale. The black histogram includes all these clean collisions, while the red histogram is restricted to systems with $\Delta{r_0} > 0.1$ AU, and the green histogram only includes MA01 stable systems (13 systems). The total (black) and initially well-separated (red) distributions are both peaked at around $10 P_{\rm in}$ and decay rapidly at longer timescales, while the MA01 stable (green) distribution is much flatter and lacks the peak completely. We note that the intersection of the red and green histograms, i.e. MA01 stable systems with $\Delta{r_0} =  a_{\rm in,0}(1 - e_{\rm in,0}) > 0.1$ AU, is almost indistinguishable from the green (MA01 stable) histogram alone, with only one less system at $< 1P_{\rm in}$. Below the top panel, we plot the cumulative fraction/number of systems resulting in these clean collisions over time.
Given that we only consider the evolution of our systems over $t_{\rm max}\sim5\times10^6 P_{\rm in}$, our estimates can be regarded as lower limits.

\begin{figure*}
\center
\includegraphics[width=\textwidth]{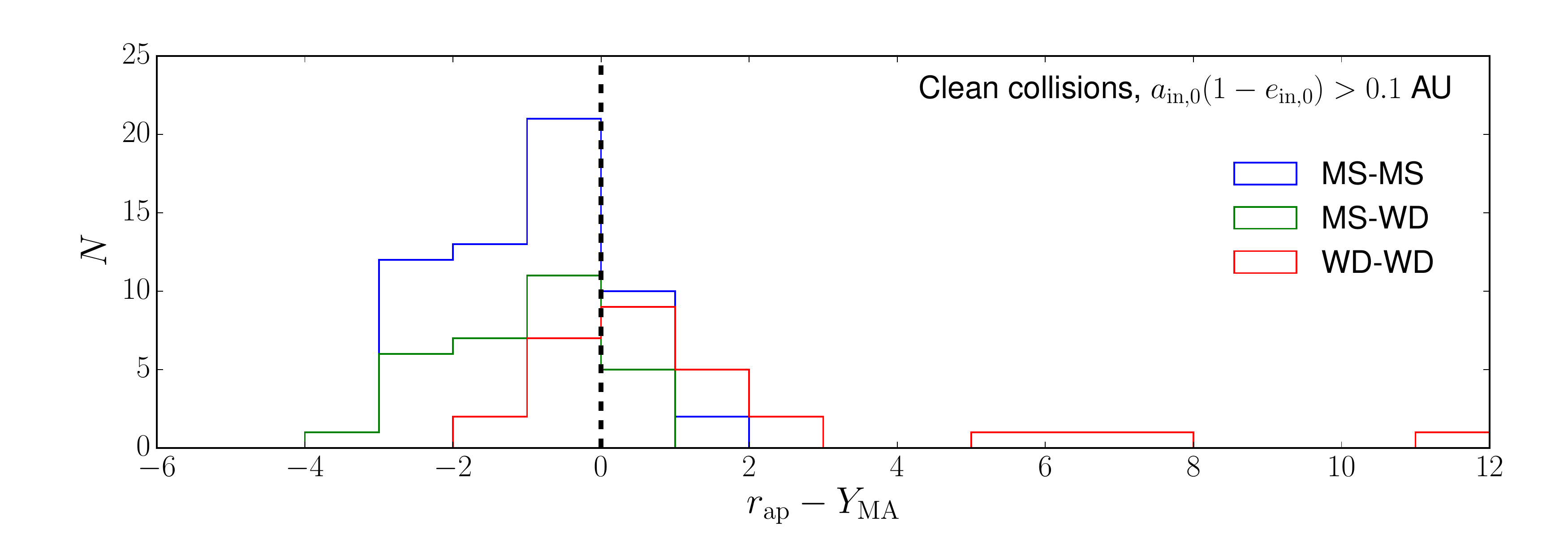}
\caption{Histograms of the \citet{MA2001} criteria parameter $f = r_{\rm ap} - Y_{\rm MA}$ for systems with clean collisions (as defined in sections 5.1 - 5.3) and $a_{\rm in,0}(1-e_{\rm in,0}) > 0.1$ AU. The majority of clean collisions involving a MS star (blue and green) involve MA01 unstable ($f < 0$) systems, while WD-WD clean collisions (red) come predominantly from MA01 stable ($f > 0$) systems.}
\label{fig:MAcleancol}
\end{figure*}

We expect clean collisions to come from either dynamically unstable configurations or 
stable ones driven by eccentricity oscillations of the inner binary due
to the torque from outer orbit.
In the latter case, based on \citet{KD2012} we expect that the torque 
from the outer orbit can overcome the dissipation barrier at $\sim6R_\odot$
and lead to a collision in one orbit if (Equation 7 therein)
\begin{align}
\frac{a_{\rm out}(1-e_{\rm out})}{a_{\rm in}} &< 2.55\left(\frac{m_3}{m_1+m_2}\right)^{1/3}
\left(\frac{a_{\rm in}}{6R_\odot}\right)^{1/6}\nonumber \\
&\simeq 6.8\left(\frac{m_3}{m_1+m_2}\right)^{1/3}
\left(\frac{a_{\rm in}}{10\mbox{ AU}}\right)^{1/6},
\label{eq:coll}
\end{align}
meaning that the clean collisions for which $e_{\rm in}\to1$ 
should typically have $r_{\rm ap}\lesssim3.6$ (2.5) if $a_{\rm in}=10$ AU
($a_{\rm in}=1$ AU). Since a rough criterion for instability 
is $r_{\rm ap}\lesssim2.5$ we expect that most clean collision
from secular torques pile up very close to the instability boundary.

In order to better quantify how close to the stability boundary the systems 
with clean collisions are, we can express the condition 
in Equation (\ref{eq:coll}) in terms of the
stability boundary parameter $Y_{\rm MA}$ from Equation (\ref{eq:MA})
as:
\begin{align}
r_{\rm ap} \lesssim{}& 2.6 ~Y_{\rm MA}
\left(\frac{a_{\rm in}}{10\mbox{ AU}}\right)^{1/6}
\frac{(m_1+m_2)^{1/15}m_3^{1/3}}{(m_1+m_2+m_3)^{2/5}}\nonumber\\
& \times\frac{(1 - e_{\rm out})^{1/5}}{(1 + e_{\rm out})^{2/5}}
  \frac{1}{1 - 0.3i_{m}/180^{\circ}},
\end{align}
which for equal-mass triple systems, 
implies that
\begin{align}
r_{\rm ap}&\lesssim 2 ~Y_{\rm MA}
\left(\frac{a_{\rm in}}{10\mbox{ AU}}\right)^{1/6},
\end{align}
so collisions indeed are expected to lie close to the stability
boundary $r_{\rm ap}=Y_{\rm MA}$.
Consistent with this expectation, from Figure \ref{fig:MAcleancol}
we observe that all the MS-MS collisions come from 
$r_{\rm ap}-Y_{\rm MA}<2$, while most do so from unstable
configurations $r_{\rm ap}-Y_{\rm MA}\lesssim0$.

\subsubsection{Collision rates in the Milky Way}

From \citet{raghavan2010} we have that $\sim18\%$ of the Solar-type stars 
are in triple stellar systems or, equivalently,  $0.09\pm0.02$ of their stellar systems are triples.
Since the ratio between the fraction of collisions and the number of stars in our simulations is $\simeq0.0125/3$, 
we derive a collision fraction of $\sim 0.18\times 0.0125\times 1/3 = 7.5\times10^{-4}$ per main sequence stars.
By taking similar numbers as in \citet{KR14}, we assume that the star formation rate in the Milky Way is constant during the last few Gyrs, and in 
steady-state we have an average collision rate of 
$\sim(10^{11}\times7.5\times10^{-4})/(10^{10})~\mbox{yr}^{-1} = 7.5\times10^{-3}$  collisions per year.

For comparison, it has been shown by \citet{KR14} that galactic tides can drive stellar collisions in a fraction of $\sim 0.1\%$ per wide binary (see Figure 2 therein). Their results indicate that very wide binaries ($a\gtrsim10^4$ AU) in our galaxy have a median probability of $\sim1/500$ of stellar collision due to galactic tides. By also assuming that $\sim5\%$ of the stars are in these wide binaries, their results imply that the fraction of main sequence stars subject to collisions is $\sim 0.05\times (1/500)\times 1/2 = 5\times 10^{-5}$. 

Our results indicate that the collision rate derived from triple systems is $\sim(7.5\times10^{-4})/(5\times10^{-5}) = 15$ times larger than that from galactic tides. Thus, contrary to the claim by \citet{KR14} that galactic tides are the dominant source of stellar collisions, we argue that triple systems dominate the rate of collisions in our galaxy. Our rate can be regarded as a lower limit for two reasons: (i) the more widely separated binaries might not have had enough time to survey the phase-space to reach the collision state, and (ii)
our definition of collision is conservative and collisions can still occur in systems that are subject to some tidal dissipation.

\subsection{Clean collisions between a main-sequence star and a
compact object}

We take advantage of the scale-free nature of the three-body dynamics
to give an estimate of the rate of a direct collision between a main sequence star and
a compact object (WD, NS, or a stellar-mass BH).
We assume that a direct collision occurs when $\Delta{r} < R_\odot$, 
while it avoids the same tidal barrier as in the previous case of two main sequence stars, $\Delta{r} \simeq 0.03$ AU, by passing through these thresholds in less than $1 P_{\rm in}$. In our simulations, this amounts to the number of systems tracked going from $10^{-1.5}$ to $10^{-2.25}$ AU in less than $1 P_{\rm in}$.

A total of 0.83\% (39/4720) of our systems involved a clean collision in this manner. Limiting our calculations to include only systems satisfying $\Delta{r_0} = a_{\rm in,0}(1 - e_{\rm in,0}) > 0.1$ AU results in 0.64\% (30/4720) as the clean collision rate of main sequence stars and compact objects. When considering only the stable systems according to the MA01 criterion, this number drops to just 0.11\% (5 systems). Thus, most of our clean collisions result in ejections if the bodies are allowed to evolve over time as point sources.

If a less conservative threshold for the tidal barrier of $\Delta r=10^{-1.75}$ AU $\simeq3.8 R_\odot$ instead of $\simeq6.7R_\odot$ is used, the number of clean collisions increases from 39 to 64.

We remark that we have ignored stellar evolution in our calculations
and it is expected to play a significant role at modifying the dynamics
in triple systems (see \citealt{toonen16} for a recent review).

\subsection{Comparison with \citet{KD2012}: rate of WD-WD collisions}

Similar to \citet{KD2012} we assume that a direct collision
between WDs happens when the distance $\Delta r<2R_{\rm WD}\sim10^{-4}$ AU, 
while avoiding the tidal barrier at $\sim2-4R_{\rm WD}$ during previous
pericentre passages. We thus consider the number of systems crossing $\Delta{r} \simeq 10^{-3.75}$ to $10^{-4}$ AU in less than one orbit, $\Delta{t} < 1 P_{\rm in}$.

We find that the number of WD-WD undergoing clean collisions is 35, while
the number of all the systems reaching $\Delta r<2R_{\rm WD}$ is 99. The number of WD-WD clean collisions occurring in systems with $\Delta{r_0} = a_{\rm in,0}(1 - e_{\rm in,0}) > 0.1$ AU is 29, equivalent to a fraction of about 0.61\% of our systems. In contrast with the main sequence star collisions (with another main sequence star or a compact object) described in the previous subsections, most of these WD-WD collisions occur in systems stable against ejections: 25 of these clean collisions occurred in MA01 stable systems. The distributions of the MA01 stability criteria $f = r_{\rm ap} - Y_{\rm MA}$ for our clean collisions are shown in Figure \ref{fig:MAcleancol}. This result is also clear from Figure \ref{fig:mergerCDF}, where the number of systems reaching $\Delta{r} \lesssim 10^{-4}$ AU in the subset of MA01 unstable systems is much fewer than the total, indicating that most of these very close encounters (a fraction of which lead to clean collisions) must come from stable systems as mentioned previously in section 4.2.

\begin{figure*}
\center
\includegraphics[width=\textwidth]{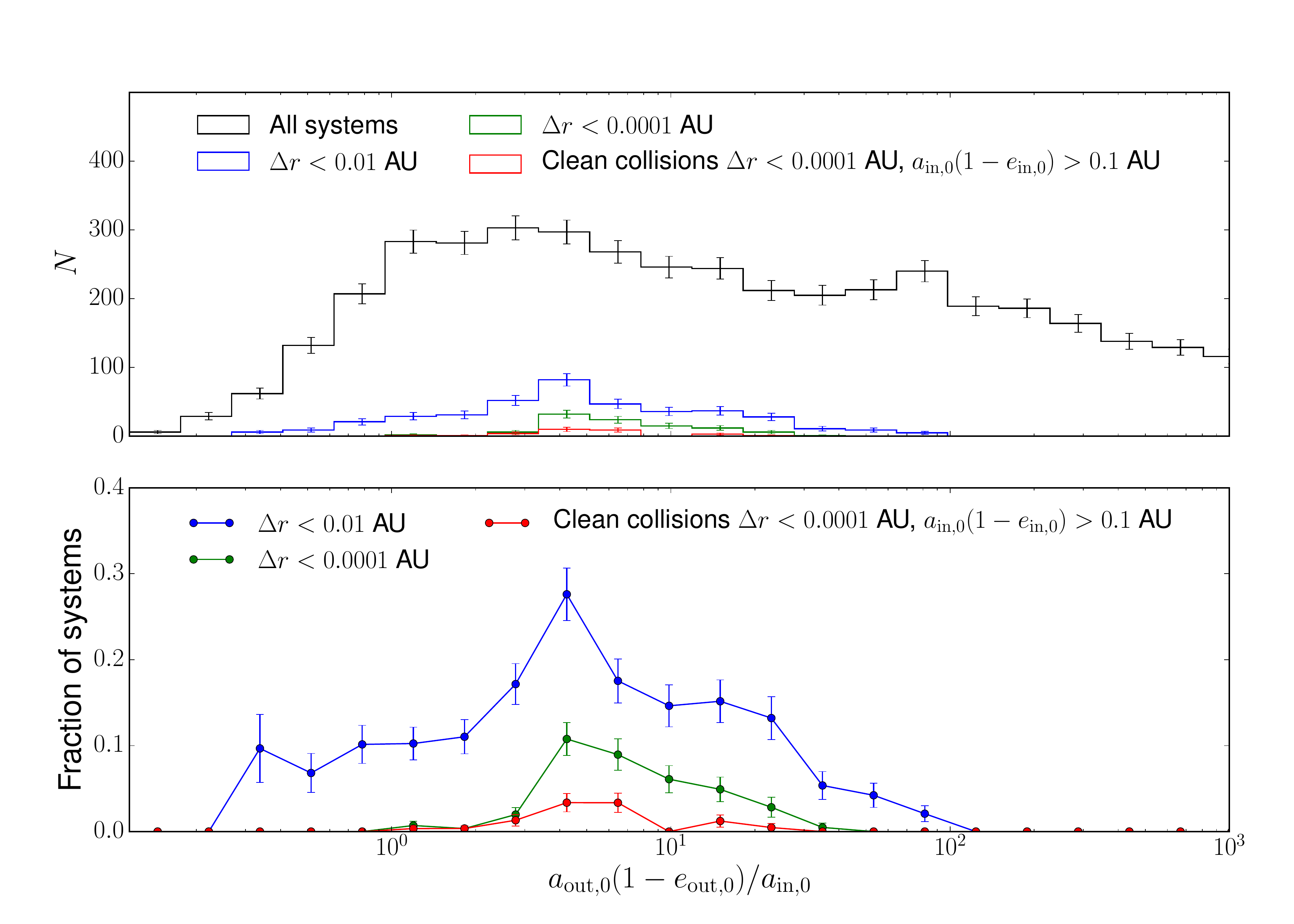}
\caption{\textbf{Top:} Histograms showing the distributions of initial outer pericentre to inner semi-major axis ratio, $a_{\rm out,0}(1-e_{\rm out,0})/a_{\rm in,0}$, for systems reaching various close encounter distances. \textbf{Bottom:} Fractions of systems reaching those close encounter thresholds as a function of $a_{\rm out,0}(1-e_{\rm out,0})/a_{\rm in,0}$. In both panels, the red curve denotes systems involving WD-WD clean collisions, as defined in section 5.3, while also initially satisfying $a_{\rm in,0}(1-e_{\rm in,0}) > 0.1$ AU. We use 50 bins distributed uniformly in log and truncate the upper end at $10^3$.}
\label{fig:mergers}
\end{figure*}

In the top panel of Figure \ref{fig:mergers}, we plot histograms showing the distributions of the ratio of initial outer pericentre to initial inner semi-major axis, $r_{\rm p,out,0}/a_{\rm in,0} = a_{\rm out,0}(1-e_{\rm out,0})/a_{\rm in,0}$, for systems reaching various close encounters ($\Delta{r} = 10^{-2}$ and $10^{-4}$ AU) and for WD-WD clean collisions (red curve). While we truncate the histogram at $r_{\rm p,out,0}/a_{\rm in,0} = 10^3$, all of the systems resulting in $\Delta{r} \leq 10^{-2}$ AU come from $r_{\rm p,out,0}/a_{\rm in,0} \lesssim 10^2$. We observe that WD-WD collisions occur preferentially in systems with $r_{\rm p,out,0}/a_{\rm in,0} \sim 3-20$.
The distribution is peaked between $r_{\rm p,out,0}/a_{\rm in,0} \sim 3-7$, with a few ($\sim3\%$) percent of these systems undergoing WD-WD clean collisions, similar to the 
results by \citet{KD2012} (Figure 3 therein).

The fraction of WD-WD collisions considering stable systems only is $29/3366\sim1\%$, which is a factor of $\sim2-5$ smaller than the 
few percent found by \citet{KD2012}. This is expected because our initial conditions span
in a wider range of separations compared to \citet{KD2012}. 
If we restrict to only stable systems with
$r_{\rm p,out,0}/a_{\rm in,0}=2-10$ as in \citet{KD2012}, we get that
$29/579\sim 5\%$ of the systems result in a direct collision in better agreement
with their results.

We note that our WD-WD collisions occur over a wide range of timescales, from after a few orbits to over $10^6$ orbits of the inner binary. It has been shown by \citet{T11} using numerical calculations that the merger time can be less than the Hubble time for a wide range of binary periods in triple systems where the tertiary induces Kozai eccentricity oscillations in the inner binary.

\subsection{Tidally-driven outcomes: short-period binaries}

%The systems reaching close approaches ($\Delta r\lesssim 20R_\odot\sim0.1$ AU)
%can lead to efficient tidal interactions, shrinking the orbits and forming
%short-period binaries.
%In particular, if we assume that all binaries reaching $\Delta r<0.1$ AU ($\Delta r<0.05$ AU)
%are tidally captured, then from Figure \ref{fig:mergerCDF} we find that $\sim29\%$
%($\sim23\%$) of our systems would form short-period binaries . 
%These number rates should be regarded as upper limits for the following reasons:
%\begin{itemize}
%\item If a binary reaches $\Delta r\lesssim 2R_\odot$ in timescales that are 
%short compared to the timescales of tidal capture and pericentre precession 
%due to short-range forces, then the stars merge instead of forming short-period 
%binaries. From Figure \ref{fig:mergerCDF}, we estimate that up to $\simeq8.5\%$ 
%could potentially lead to mergers or disruptions;
%\item some systems might not spend enough time at small distances, so
%tides might not be efficient at shrinking the binary orbits within the age of the 
%systems.
%\end{itemize}

%Given these arguments, the actual rate of formation of short-period binaries 
%depends critically on the details of tides, which we have not included in
%our integrations.
%However, we are still able to constrain the rate of different possible
%outcomes of the systems reaching close approaches because these depend
%most critically on the closest approaches between bodies and how fast they
%reach these distances.

Even though we have neither included tidal dissipation nor 
tidal disruptions in our calculations, we can estimate a good 
upper limit to the fraction of
short-period binaries that can be formed. 
We shall define 
short-period binaries as the systems where the period
of the inner binary is
$\lesssim10$ days or for the purpose of our work
with final semi-major axis $a<0.1$ AU.

We note that short-period binaries can be primordial due to our
choice of initial conditions or due to eccentricity excitation
from the tertiary
followed by tidal circularization (e.g., \citealt{MS79,FT2007}).

First, we estimate the former fraction of primordial binaries
for which the tertiary plays no role in its formation.
We notice that these can come from initially circular binaries 
with $a<0.1$ AU or eccentric binaries with any value $a$
followed by circularization.
As in previous studies (e.g., \citealt{EK2006,FT2007}) we estimate
the timescale to shrink the orbit from equilibrium tides for 
two Solar-type stars as (e.g., Equation 4 in \citealt{petrovich2015b}) 
\begin{align}
\tau_a \equiv{}&
\left(\frac{1}{a\in}\frac{da\in}{dt}\right)^{-1}
\sim 1\mbox{ Gyr~}\left(\frac{t_{\rm V}}{5~\mbox{yr}}\right)
\frac{1}{f(e\in)}\nonumber\\
&\times \left(\frac{a\in[1-e\in^2]}{0.1 \mbox{ AU}}\right)^{15/2}
\left(\frac{a\in}{1 \mbox{ AU}}\right)^{1/2},
\end{align}
where $f(e)=e^2(1+15/4e^2+15/8e^5+5/64e^6)$
and $t_V$ is the viscous time of the stars.
Thus, the systems where the inner binary 
has $a[1-e^2]\lesssim0.1$ AU can form
a short-period binary in the main sequence's lifetime of the star.
A fraction of $136/3193 \sim 4\%$ of our simulated systems 
satisfy this condition.

Second, some binaries starting with $a[1-e^2]>0.1$ AU
(not a primordial short-period binary)
can acquire high eccentricities
driven by the tertiary and reach $a[1-e^2]\lesssim0.1$ AU.
In the limit of high eccentricities ($e\gtrsim0.9$)
the condition $a[1-e^2]\lesssim0.1$ AU reads 
$a[1-e]\lesssim0.05$ AU.
In Figure \ref{fig:mergerCDF} we show the minimum pericentre distances (black circles) and 
find that $\sim23\%$ of them reach $a[1-e]\lesssim0.05$ AU at some
point of their evolution, while $\sim10\%$ do so 
from the beginning of the integration (blue crosses).

Since about half of our short-period binaries 
are formed because of our initial conditions (initial 
$a\in[1-e\in^2]\lesssim0.1$ AU), up to $\sim 13\%$ of all our systems are 
formed by a high-eccentricity path ($a[1-e]\lesssim0.05$ AU)
due to the gravitational perturbation from the third body.
This is an upper limit as some systems will lead to mergers
(roughly $\sim1\%$), while others will not circularize because
the inner orbit spends little time at 
the minimum $\Delta r=a\in(1-e\in)$.

Our results are roughly consistent with those from 
\citet{NF2014} who found that $\sim21\%$ 
of the triple systems have a short-period binary at the end
of their integration with
only $\sim40\%$ of these ($\sim8\%$ of the triple systems) being 
brought into their short-period orbits due to the gravitational
perturbations from the tertiary.
We note that the authors used the secular equations of motion with tidal
dissipation and systems initially in MA01 stable orbits for which
$\epsilon_{\rm oct}=a\in e\out/[a\out(1-e\out^2)]<0.1$.
If we restrict to only those systems we get a similar result
for the formation of short-period binaries and 
we confirm this result by integrating the systems with the
secular code in \citet{petrovich2015b} including relativistic
precession but no tidal interactions.

\section{Summary}
In this work, we run a set of N-body integrations using a wide range of initial conditions for stellar triple systems to study their stability and dynamical outcomes. 
The main goals of our simulations are
to test the previously proposed and commonly relied-upon stability criteria in 
the literature, namely by \citet{EK1995} and \citet{MA2001} specifically for these 
types of triple systems, 
and to determine the rate at which triple-stellar systems merge
in the galactic field.

Our main results are summarized below:

\begin{itemize}
\item The EK95 and MA01 criteria perform well in characterizing the stability of these systems against ejections over timescales of $5\times10^6 P_{\rm in}$ for a wide distribution of orbital configurations for stellar triples motivated by the observed populations in \citet{DM1991}. 
The MA01 criteria performs slightly better when considering the completeness of both stable and unstable classifications altogether, while EK95 has a similar performance as the simple
criterion $r_{\rm ap}\equiv a\out(1-e\out)/[a\in(1+e\in)]=2.5$.
We recommend the usage of the MA01 criterion for classifying systems as either 
long-term stable or unstable based on their initial conditions.

\item We estimate the rate of stellar collisions between main sequence stars, compact objects, and combinations of these stars arising from dynamical effects which drive the eccentricities of their orbits up to cause close pericentre passages and other close encounters between the bodies. 
Even with a rather strict definition of a clean collision\footnote{A clean collision is defined as those in which the bodies pass through their mutual tidal barrier in less than one orbit to ensure that tidal forces would not circularize their orbits, which we did not directly simulate.} at these various scales , and only considering initially well-separated systems ($\Delta{r_0} = a_{\rm in,0}(1 - e_{\rm in,0}) > 0.1$ AU), we still get non-negligible fractions of systems which result in clean collisions: $\sim$ 1.3\% (MS-MS), 0.64\% (MS-Compact Object), and 0.6\% (WD-WD). These values are regarded as lower limits and imply significant rates of head-on stellar collisions resulting from triple interactions.

\item We find that the rate of MS-MS collisions from triple stellar systems 
is $\sim7\times10^{-3}~\mbox{yr}^{-1}$, 
which is  at least 15 times higher than the rate derived 
for very wide binary systems affected by galactic tides found by \citet{KR14}. 
Also, \citet{PK2012} derived a rate of stellar collisions resulting from the dynamical instability of triple systems caused by mass loss, which they term the triple evolution dynamical instability (TEDI), of $\sim 1.2 \times 10^{-4} ~\mbox{yr}^{-1}$. 
Thus, we conclude that in addition to the stellar collision rate derived from their TEDI process, the orbital evolution of the triple systems alone provides enough clean collisions to dominate the stellar collision rate in the Milky Way.

\item By tracking the closest approaches between stars, but not
including tides in our simulations, 
we estimate an upper limit to the fraction of short-period binaries in
triple systems of $\sim23\%$. Only $\sim58\%$ of these ($\sim 13\%$ of the
systems) are due to triple interactions, results that are consistent
with those from \citet{NF2014} using a secular code that includes tidal
interactions.
\end{itemize}

Finally, we remark that although we only considered the dynamical evolution of our triple systems in our simulations, it has been shown that stellar evolution can also play an important role and affect the outcomes of these systems. As previously mentioned, \citet{PK2012} considered the effects of mass loss of the primary and argue that the ratio $a_{\rm out}/a_{\rm in}$ decreases since $a_{\rm in}$ increases more than $a_{\rm out}$, due to the larger relative mass loss in the inner system compared to the outer system. Thus, dynamically stable systems can move into the unstable regime due to mass loss, a process \citet{PK2012} called the TEDI mechanism. Furthermore, if the primary becomes much less massive than the other two stars, as may be the case if the primary suffers severe mass loss in becoming a WD, the system may become subject to the Eccentric Kozai-Lidov mechanism and develop instability (the Mass-loss Induced Eccentric Kozai mechanism, \citealt{ST13}). On the other hand, triple systems can also become more stable if there is significant mass loss/transfer of the third, outer companion, which may quench secular evolution (secular evolution freeze-out, \citealt{MP14}). These studies reveal that the dynamical outcomes of triple systems can be influenced by many factors of their stellar evolution, and our understanding of these outcomes will benefit from additional future studies involving both orbital and stellar evolution.

\section*{Acknowledgements}
% Entry for the table of contents, for this guide only
\addcontentsline{toc}{section}{Acknowledgements}
This project is the result of the 2016 Summer Undergraduate Research Program (SURP) in astronomy \& astrophysics at the University of Toronto. 
The integrations in this paper made use of the REBOUND code which can be downloaded freely at \mbox{\url{http://github.com/hannorein/rebound}}.
This research was made possible by the computing resources at the Canadian Institute for Theoretical Astrophysics.
We thank Rebekah Dawson for carefully reading our manuscript and providing detailed comments.
We also thank Daniel Tamayo, Ondrej Pejcha, and Yanqin Wu
for useful discussions.
C.P. is grateful for support from the Jeffrey L. Bishop Fellowship
and from the Gruber Foundation Fellowship.

%%%%%%%%%%%%%%%%%%%%%%%%%%%%%%%%%%%%%%%%%%%%%%%%%%



%\bibliographystyle{mnras}


\bibliographystyle{mn2e}
%\bibliography{refs_ryan.bib}
%\bibliography{../1Mybib.bib}

%%%%%%%%%%%%%%%%%%%%%%%%%%%%%%%%%%%%%%%%%%%%%%%%%%

%%%%%%%%%%%%%%%%% APPENDICES %%%%%%%%%%%%%%%%%%%%%

%\appendix

%%%%%%%%%%%%%%%%%%%%%%%%%%%%%%%%%%%%%%%%%%%%%%%%%%

% Don't change these lines
\bsp	% typesetting comment
\label{lastpage}
\end{document}